\documentclass[twocolumn,a4paper, traditabstract]{aa}

\usepackage{amsmath,amssymb}

\usepackage{graphicx,natbib,verbatim,caption}
\usepackage[varg]{txfonts}

\bibpunct{(}{)}{;}{a}{}{,}

\begin{document}

\titlerunning{The temperature and density structures of coronal X-ray sources}
\authorrunning{Jeffrey, Kontar \& Dennis}

\title{High-temperature differential emission measure and altitude variations in the temperature and density of solar flare coronal X-ray sources}

\author{Natasha L. S. Jeffrey\inst{1}, Eduard P. Kontar\inst{1} \and Brian R. Dennis\inst{2}}

\offprints{{N. L. S. Jeffrey \email{natasha.jeffrey@glasgow.ac.uk}}}

\institute{School of Physics \& Astronomy, University of Glasgow, Glasgow G12 8QQ, UK \and Solar Physics Laboratory, Code 671, Heliophysics Science Division, NASA Goddard Space Flight Center, Greenbelt, MD, USA, 20771}

\date{Received ; Accepted}

\abstract{The detailed knowledge of plasma heating and acceleration region properties
presents a major observational challenge in solar flare physics.
Using the {\em Ramaty High Energy Solar Spectroscopic Imager (RHESSI)},
the high temperature differential emission measure, $DEM(T)$, and the energy-dependent spatial
structure of solar flare coronal sources are studied quantitatively. 
The altitude of the coronal X-ray source is observed to increase with energy
by $\sim +0.2$ arcsec/keV between 10 and 25 keV.
Although an isothermal model can fit the thermal
X-ray spectrum observed by {\em RHESSI}, such a model {\it cannot}
account for the changes in altitude, and multi-thermal coronal sources are required where the temperature increases with altitude. For the first time, we show how {\em RHESSI} imaging information can be used to constrain the $DEM(T)$ of a flaring plasma.
We develop a thermal bremsstrahlung X-ray emission model with inhomogeneous temperature
and density distributions to simultaneously reproduce: i) $DEM(T)$,
ii) altitude as a function of energy, and iii) vertical extent of the flaring coronal
source versus energy. We find that the temperature-altitude gradient in the region is
$\sim+0.08$ keV/arcsec ($\sim$~1.3~MK/Mm).
Similar altitude-energy trends in other flares suggest that the majority of coronal X-ray sources
are multi-thermal and have strong vertical temperature and density gradients with a broad $DEM(T)$.}


\maketitle

\section{Introduction}\label{intro}

Over the last decade, the X-ray imaging spectroscopy of the {\it Ramaty High Energy Solar
Spectroscopic Imager (RHESSI)} \citep{2002SoPh..210....3L} has allowed changes in the spatial properties
of solar flare X-ray sources to be studied in detail. As well as high energy resolution X-ray spectroscopy
($\le$1 keV at 3 keV increasing to 5 keV at 5 MeV), {\em RHESSI} is capable of non-direct X-ray imaging using nine rotating
modulation collimators (RMCs) giving angular resolutions between $2\arcsec.3$ and $183\arcsec$.
In practice, the angular resolution is usually $\gtrsim5\arcsec$ (Full Width Half Maximum) due to finite counting statistics and image reconstruction uncertainties.
However, {\em RHESSI} is capable of inferring changes in the positions of X-ray
sources down to the sub-arcsecond level.
Sub-arcsecond measurements of X-ray footpoint locations have been achieved using X-ray visibilities \cite[see e.g. ][]{2008A&A...489L..57K,2010ApJ...717..250K,2013ApJ...766...75J}, improving
upon forward fitting a Gaussian source model to the {\em RHESSI}
modulated lightcurves \citep[e.g.][]{2002SoPh..210..383A}.

Recently, many studies have investigated flares where the majority of X-ray emission
comes from a coronal thick-target source \citep[e.g.][]{2004ApJ...603L.117V}, in contrast to the usual flare that is dominated
by footpoint hard X-ray (HXR) emission. The first energy dependent spatial study of such coronal X-ray sources,
\citep{2008ApJ...673..576X} and further works \citep[e.g.][]{2012ApJ...755...32G,2013ApJ...766...28G}
examined how X-ray source lengths (the direction that appears to lie parallel to a guiding magnetic field)
changed with X-ray energy. Such observations, with the help of numerical simulation \citep{2014ApJ...787...86J},
have allowed the estimation of coronal plasma number density and the acceleration rate of electrons within the region.
\citet{2011ApJ...730L..22K} found increasing X-ray source widths (defined in the direction perpendicular
to the guiding field) with energy, a trend consistent with magnetic turbulence in the flaring coronal source.
\cite{2013ApJ...766...75J} studied the temporal evolution of radial positions (altitudes),
lengths and widths of such coronal X-ray sources.

Changes in coronal X-ray source radial position (or altitude) with X-ray energy have not been studied extensively in the {\em RHESSI} era. Unlike the changes in X-ray source size,
an increasing altitude with time is often observed for types of
flaring coronal X-ray sources, and the trend is often seen in other
wavelengths such as UV, EUV and soft X-rays \citep[e.g.][]{1996ApJ...459..330F}.
In a standard flare model, the upward motion of coronal X-ray sources with time is often explained by the upward motion of a magnetic reconnection site,
with loops reconnecting continuously  at increasing altitude and then cooling \citep[e.g.][]{1987SoPh..108..237S,1992PASJ...44L..63T,1996SoPh..169..403S,2002SoPh..210..341G}.
Sometimes an initial decrease in altitude is followed by an increase in altitude after the impulsive phase of the flare  \citep{2003ApJ...596L.251S,2004ApJ...612..546S,2006A&A...446..675V,2009ApJ...696..121L,2009ApJ...706.1438J,2010ApJ...724..171R,2012ApJ...749...85G,2013ApJ...766...75J},
and sometimes even more complicated motions are observed \citep[e.g.][]{2013ApJ...767..168L}.
These observations are  viewed as an argument to support the standard
flare scenario involving magnetic reconnection above the coronal source.
Despite {\it RHESSI's\,} unprecedented spectral resolution, the temperature structure (differential emission measure $DEM(T)$) above $T\sim1$ keV remains poorly determined \citep[e.g.][]{2006SoPh..237...61P}.

In this paper, we present a spatial-spectral study of flaring coronal X-ray sources.
A limb flare SOL2013-05-13T02:12 is studied in detail (the time of 02:12 UT is the {\em RHESSI\,} peak flux time).
We deduce the changes in coronal X-ray source location with energy at a given time and find a relationship between X-ray source energy
and altitude. During this study, we are not concerned with the physical cause of the trend, only how {\em RHESSI} imaging information can be used to further constrain the $DEM(T)$, so that it is consistent with {\it both} {\em RHESSI} spectral {\it and} imaging observations. Further, an isothermal model, often used to fit the X-ray spectrum at low energies,
is shown to be inconsistent with {\em RHESSI} imaging observations for the flares studied.

\section{The observation of SOL2013-05-13T02:12}\label{obs_intro}

\begin{figure}
\centering
\captionsetup{width=1.0\linewidth}
\includegraphics[width=10cm]{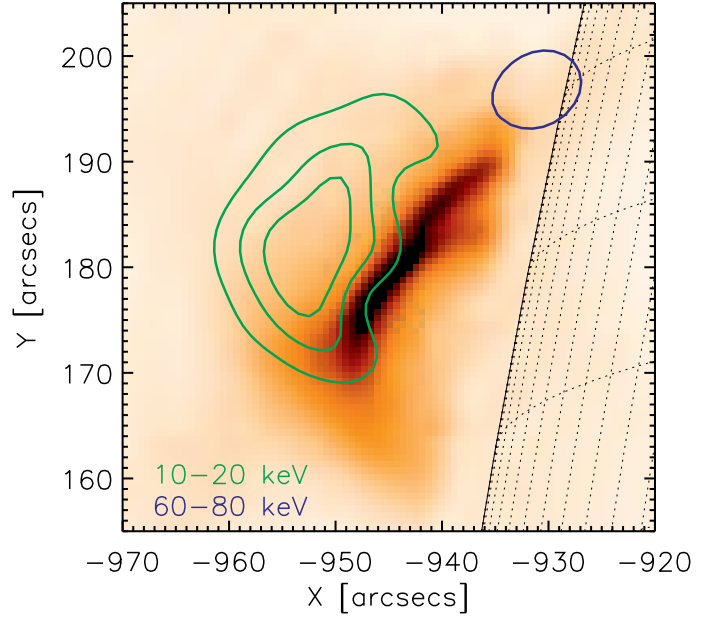}
\caption{{\em SDO} AIA 94~\AA~image at 02:08:40 UT (red-orange background). {\em RHESSI} CLEAN contours at the time interval of 02:08 to 02:10 UT for 10-20 keV (green) at 30, 50 and 70\% of the maximum, and for 60-80 keV (blue) at the 50 \% level.}
\label{flare_intro}
\end{figure}

\begin{figure}
\centering
\captionsetup{width=1.0\linewidth}\hspace{-1.0cm}
\includegraphics[width=9.5cm]{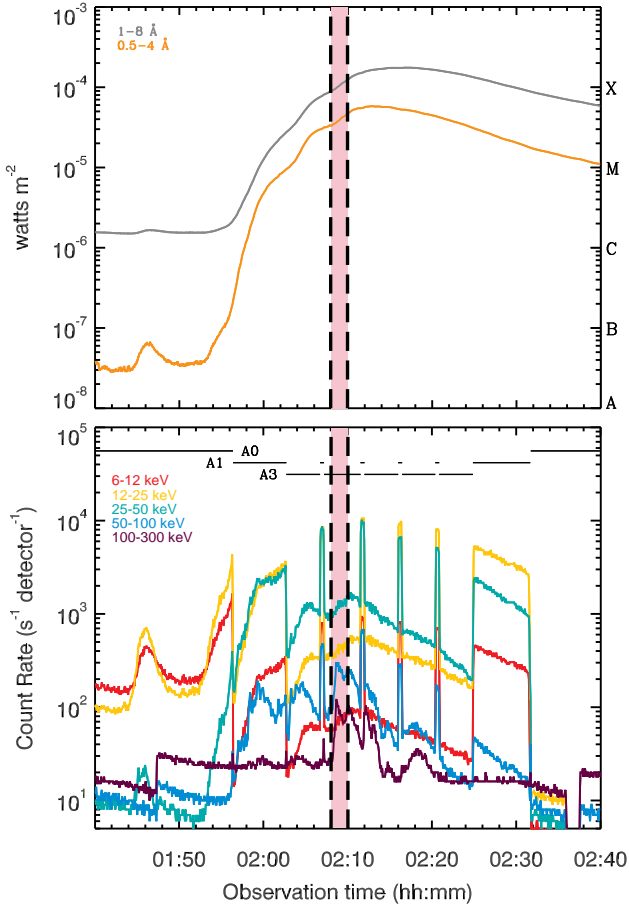}
\caption{{\em GOES\,} (top: 1-8\;\AA\;and 0.5-4\;\AA) and {\em RHESSI} (bottom: 6-12, 12-25, 25-50, 50-100 and 100-300 keV) lightcurves for the chosen flare SOL2013-05-13T02:12. The two minute time of study from 02:08 to 02:10 UT is shown by the pink band between the two vertical dashed lines. The `jumps' in the {\em RHESSI} lightcurve occur due to instrumental attenuation changes reducing the X-ray flux reaching the detectors. At this time, {\em RHESSI} was in attenuator state A3, meaning that both the thin and thick attenuators cover the detectors.}
\label{lightcurve}
\end{figure}

GOES X1.7 flare SOL2013-05-13T02:12 was chosen for detailed analysis because it had a strong coronal X-ray
source located at the eastern limb.
A limb flare was chosen so that the radial distance changes
correspond to height changes with minimal projection effects.
This flare has one visible northern footpoint that can be seen
up to energies of $\sim100$~keV. Figure \ref{flare_intro} shows a loop filled with hot plasma imaged with
the {\em Solar Dynamics Observatory Atmospheric Imaging Assembly} ({\em SDO} AIA) \citep{2012SoPh..275...17L} in the 94~\AA~passband. {\em RHESSI} 10-20 keV and 60-80 keV contours are displayed.
The 10-20 keV emission comes from the flare coronal source
located above the 94~\AA~loop, while the bulk of the 60-80 keV emission comes from a northern footpoint source located close
to the footpoint of the 94~\AA~loop. The {\em RHESSI} and {\it GOES\,} lightcurves for the flare are shown
in Figure \ref{lightcurve}.
There is a gradual rise in the X-ray flux up to 50 keV from 01:56 UT onwards,
with sharper increases in the 50-100 keV and 100-300 keV bands between 01:56 and 02:30 UT.

\section{{\em RHESSI} spectroscopy and imaging}\label{rhes_spec}

\begin{figure*}
\centering
\includegraphics[width=6.6cm]{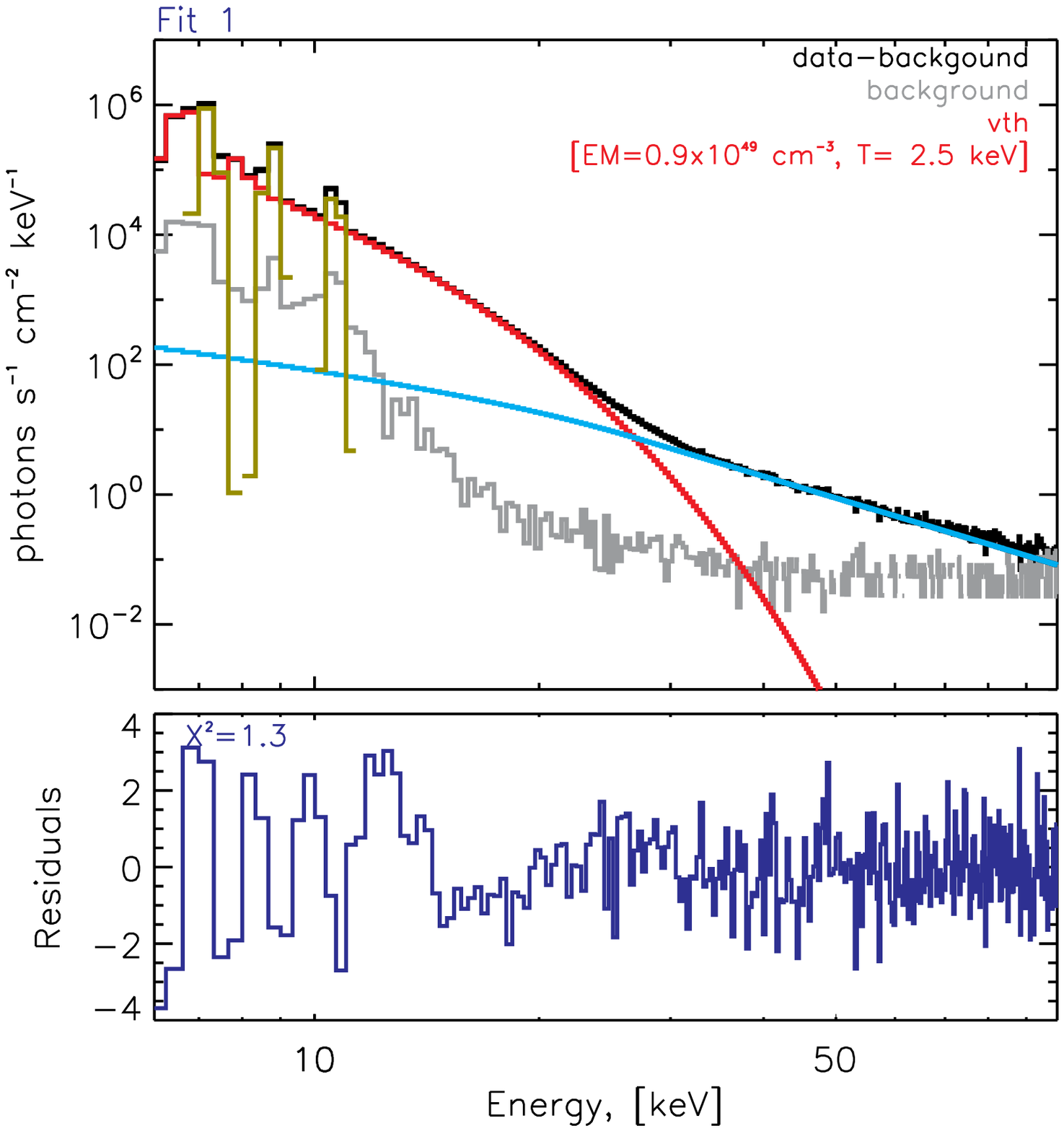}
\hspace{-1.5cm}
\includegraphics[width=6.6cm]{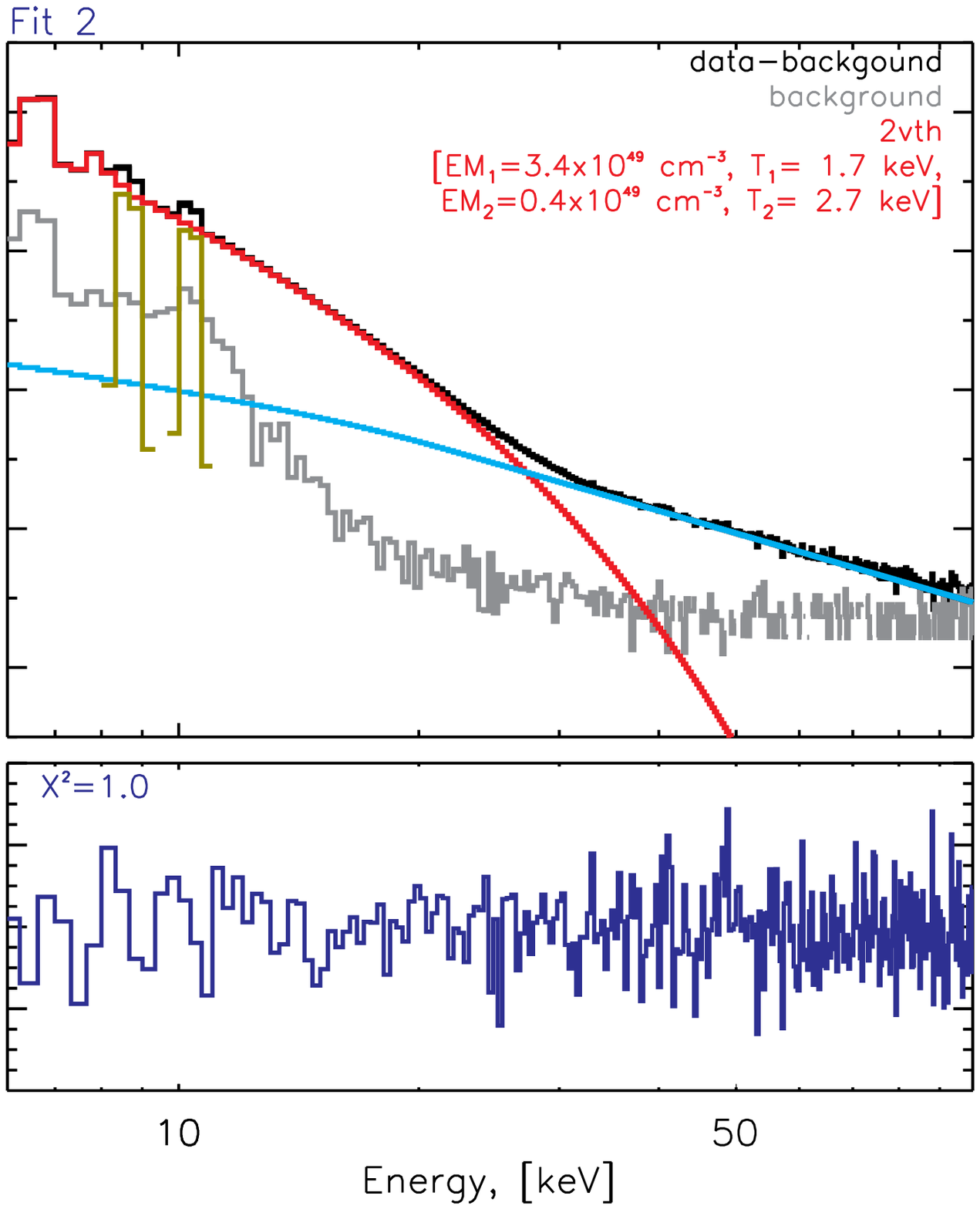}
\hspace{1.5cm}
\includegraphics[width=6.6cm]{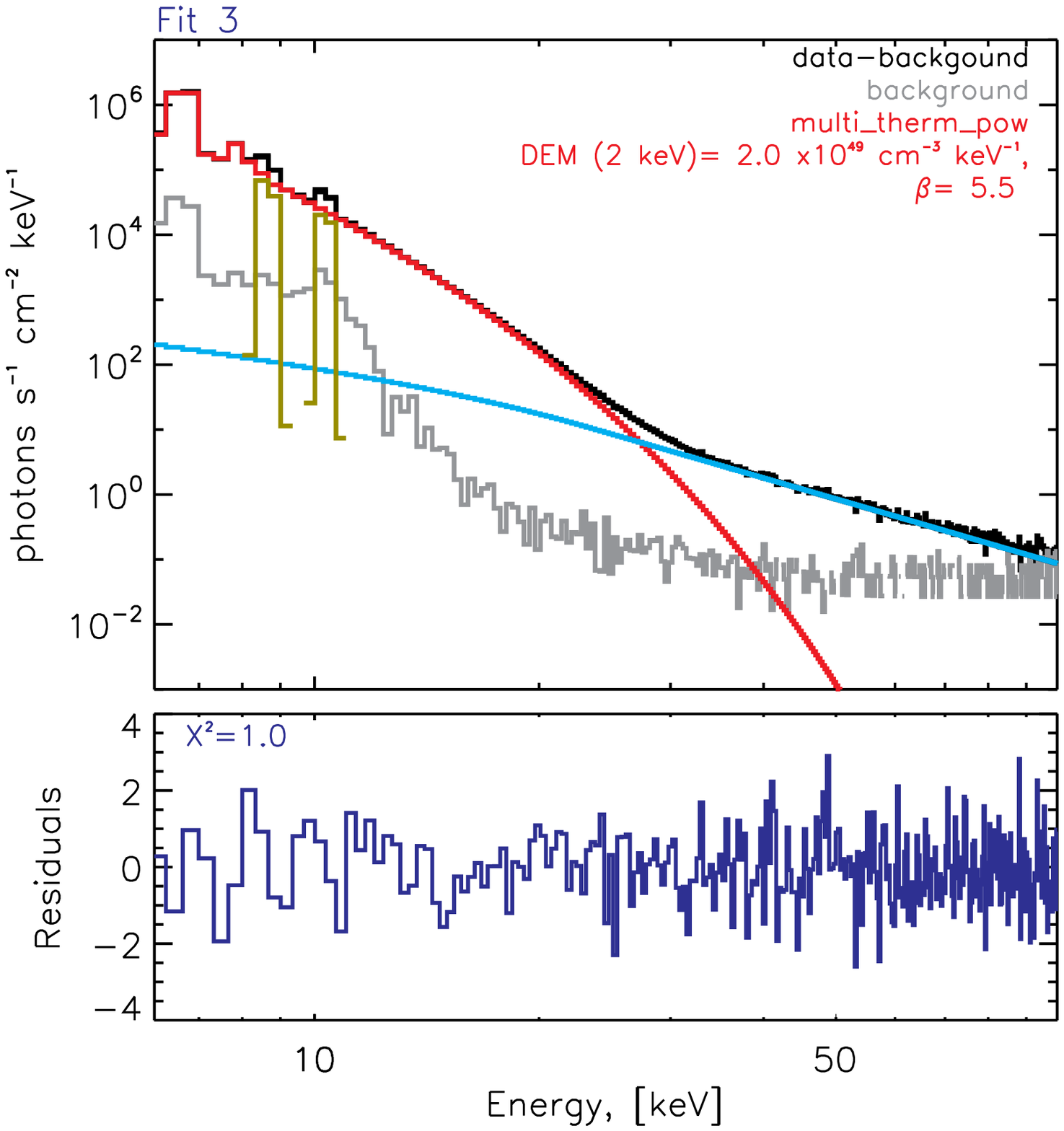}
\hspace{-1.7cm}
\includegraphics[width=6.6cm]{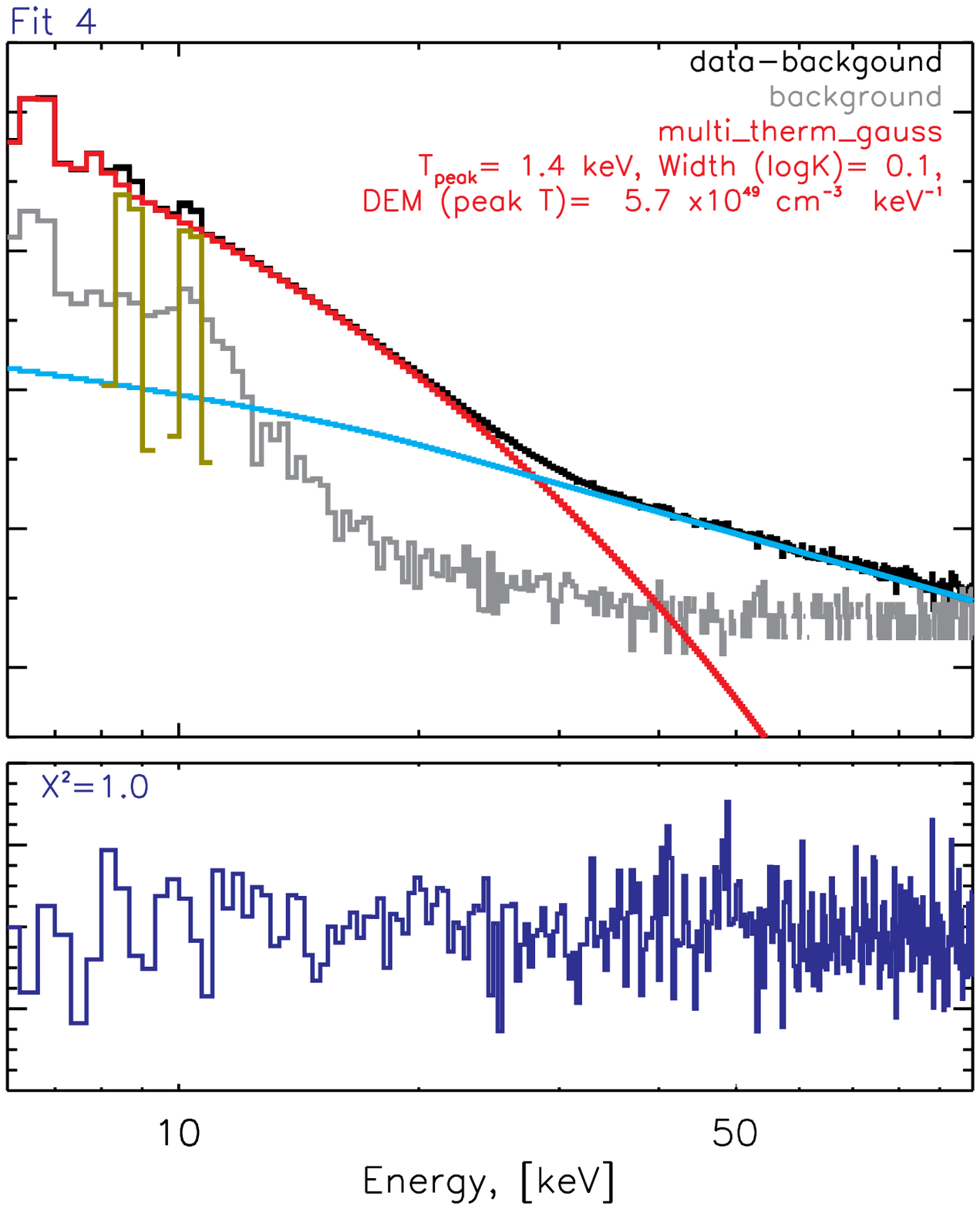}
\hspace{-1.7cm}
\includegraphics[width=6.6cm]{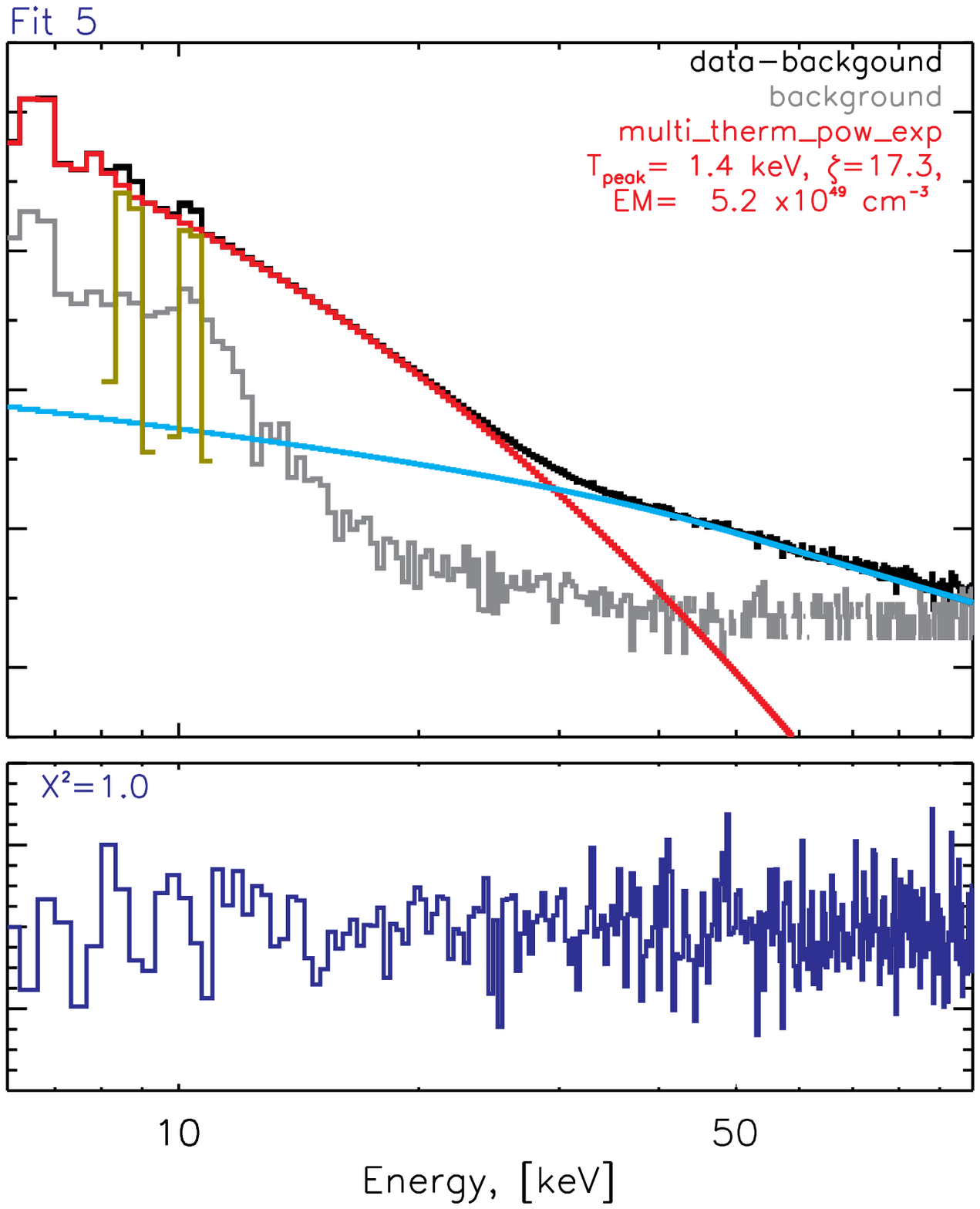}
\caption{The photon flux spectrum fitted with different functions describing the thermal component.
{\it Top left:} single isothermal, {\it Top right:} double isothermal, {\it Bottom left:} multi-thermal power law.
{\it Bottom middle:} multi-thermal Gaussian in $\log_{10}T$ and
{\it Bottom right:} multi-thermal power-law and exponential.
The normalised residuals are plotted below each spectrum, and each is created using only detector 6 and the functions fit the count spectrum
between 6 and 100 keV for a chosen time interval of 02:08 to 02:10 UT. The values of all thermal parameters are shown in the figure and in Table \ref{thermal_paras}. The gold lines in each panel represent Gaussian line fits (line) compensating for instrumental anomalies in the {\em RHESSI} spectrum. A thick2\_vnorm function (light blue) is used to account for the non-thermal X-ray emission at higher energies.}
\label{fig:spectra}
\end{figure*}

A spectral analysis of the flare was performed for the time interval from 02:08 to 02:10 UT.
The count rate for this X-class flare is high so spectroscopy using a single detector could be performed.
Comparison of  individual spectra from detectors 1 to 9 showed
that detector 6 had the best energy resolution, and shows the spectral features between 6 to 10 keV.
As expected, the spectrum during this time interval has a strong thermal component below 30 keV
and a power-law spectrum at higher energies up to $\sim$150 keV (see Figure \ref{fig:spectra}).
Since {\em RHESSI} is in attenuator state A3  (the thin and thick attenuators cover the detectors),
the lower energy limit was set to 6 keV. The majority of the counts recorded below 6 keV
are not incident photons at that energy; they are from high energy photons producing K-shell escape
photons from the germanium detector itself \citep[see e.g.][]{2006ApJ...647.1480P}.

Using the Object Spectral Executive \citep[OSPEX software,][]{2002SoPh..210..165S}, the following five functions
describing the thermal component were fitted to the background-subtracted count fluxes
in the energy range between 6 and 100 keV:
\begin{enumerate}
\renewcommand*\labelenumi{(\theenumi)}
\item The isothermal function (f\_vth), `Fit 1', provides the temperature $T$ [keV] and emission measure $EM$ [cm$^{-3}$] of the thermal source. The emission measure and temperature are free parameters while the relative iron abundance is fixed at 1.0 times the CHIANTI atomic database (\cite{1997A&AS..125..149D,2013ApJ...763...86L}) coronal value, which is the default OSPEX value (investigating different relative iron abundances is beyond the scope of the paper). The relative iron abundance is fixed at this value for all fits (1) to (5) (see  Table \ref{thermal_paras} and Figure \ref{fig:spectra} for all parameters and fits).

\item The double isothermal function (f\_2vth), `Fit 2', is the sum of two isothermal functions each with their own emission measures
($EM_1$, $EM_2$ ) and temperatures ($T_1$, $T_2$), but with the same, fixed relative iron abundance. $EM_1$, $EM_2$, $T_1$, $T_2$ are all free parameters.

\item  A multi-thermal power-law function (f\_multi\_therm\_pow), `Fit 3', relating the differential emission measure $DEM(T)$ [cm$^{-3}$ keV$^{-1}$] to the temperature $T$ by,
\begin{equation}\label{spect1}
DEM(T)=DEM({\rm T=2\; keV})\left(\frac{2}{T}\right)^{\beta}\;,
\end{equation}
where $\beta$ is the power-law index and $DEM({\rm 2\; keV})$ is the differential emission measure at a temperature of 2 keV. The function also provides a minimum and maximum value of temperature $T$. The relative iron abundance is fixed, while all other parameters are free.

\item  A multi-thermal Gaussian in $\log_{10}T$ (f\_multi\_therm\_gauss), `Fit 4', relates the $DEM$ to the temperature $T$ using,
\begin{equation}\label{spect_g}
DEM(T)=DEM(T_{peak})\exp{\left(-\frac{(\log_{10} T-\log_{10} T_{peak})^2}{2\sigma^{2}}\right)}\;,
\end{equation}
where the $DEM$ is a Gaussian in logarithmic temperature space, $T_{peak}$ [keV] is the temperature at the peak, $DEM(T_{peak})$ is the $DEM$ at $T_{peak}$ and $\sigma$ is the standard deviation of the Gaussian in units of $\log_{10}$~keV. The relative iron abundance, and the minimum and maximum temperatures, are fixed. $DEM(T_{peak})$, $T_{peak}$ and $\sigma$ are free parameters.

\item Finally a multi-thermal power-law and exponential function (f\_multi\_therm\_pow\_exp), `Fit 5' relates the $DEM$ to the temperature using,
\begin{equation}\label{spect_mpe}
DEM(T)=\frac{EM}{\zeta T_{peak}\Gamma(\zeta-1)}\left(\frac{\zeta T_{peak}}{T}\right)^\zeta \exp \left(-\frac{\zeta T_{peak}}{T}\right),
\end{equation}
where $\Gamma(x)$ is the gamma function, $EM$ [cm$^{-3}$] is the total emission measure integrated from the minimum to the maximum temperature, $T_{peak}$ is the peak temperature of the $DEM$ and $\zeta$ is the power-law index. $EM$, $T_{peak}$ and $\zeta$ are free parameters while the relative iron abundance, and minimum and maximum temperatures, are fixed.
\end{enumerate}

The following functions were also included in each case (see Figure \ref{fig:spectra}):
thick-target bremsstrahlung (f\_thick2\_vnorm) to account for non-thermal emission from a power-law distribution of electrons above $\sim 30$~keV,
a pileup correction\footnote{ (f\_pileup\_mod) The pile-up component accounts for those photons arriving at the detector
at nearly the same time, that are detected as a single count
with an energy equal to the sum of the individual photon energies.},
a function to fine tune the {\em RHESSI} detector response matrix (f\_drm\_mod),
and line components (f\_line) to account for instrumental features near $\sim10$ keV.

Our spectral fit results show that it is impossible to constrain the overall shape of the $DEM(T)$ below $\sim T=1.5$ keV with {\em RHESSI} data. The isothermal function \citep[which is commonly used in {\em RHESSI} spectral analysis, see e.g. ][as reviews]{2011SSRv..159..107H,2011SSRv..159..301K} is a marginally worse fit in terms of the reduced $\chi ^2$  and the residuals below 15 keV (Table \ref{thermal_paras} and Figure \ref{fig:spectra}).
Otherwise, all thermal models can adequately fit the thermal part of the spectrum. This result clearly demonstrates that a wide range of temperature
distributions (summarised in Figure \ref{fig:dems}) with various $DEM(T)$ are consistent with the measured {\em RHESSI} count flux spectrum. In Figure \ref{fig:dems}, the $DEMs$ for Fit 3 (grey), Fit 4 (orange) and Fit 5 (blue) are shown. Over the temperature range of $\sim$ 1.6 keV to 3.0 keV, the OSPEX model $DEMs$ agree within the model errors (shaded areas), suggesting that {\em RHESSI} can constrain the form of the $DEM(T)$ between this temperature range, regardless of the chosen $DEM$ model. The shaded areas are derived from the formal uncertainties on each free parameter in the models determined assuming Poisson statistics are the only source of error and that all are orthogonal. A more detailed Monte Carlo analysis would be needed to determine these uncertainties more accurately using the methods described in \citet{2013ApJ...769...89I}, but the range of applicability of each model shown in Figure \ref{fig:dems} is adequate for our current purposes.

\begin{table*}[t]
\begin{center}
     \caption{A parameter comparison of the thermal fitting functions shown in Figure \ref{fig:spectra}.}
    \begin{tabular}{c c c c c c c c c c}
    \hline\hline
    Fit function & $T_{1}$ & $T_{2}$ & $T_{peak}$ & $EM_{1}$ $\times10^{49}$ & $EM_{2}$ $\times10^{49}$ & $DEM$ (2 keV) $\times10^{49}$ & $DEM$ ($T_{peak}$) $\times10^{49}$ & $\chi^{2}$ \\
     & [keV] & [keV] & [keV] & [cm$^{-3}$] & [cm$^{-3}$] & [keV$^{-1}$ cm$^{-3}$] & [keV$^{-1}$ cm$^{-3}$] & \\ \hline
    vth & 2.5 & X & X & 0.9 & X & X & X & 1.3 \\ \hline
    2vth & 1.7 & 2.7 & X & 3.4 & 0.4 & X & X & 1.0 \\ \hline
    multi\_therm\_pow & X & X & X & X & X & 2.0 & X & 1.0 \\ \hline
    multi\_therm\_gauss & X & X & 1.4 & X & X & X & 5.7 & 1.0 \\ \hline
    multi\_therm\_pow\_exp & X & X & 1.4 & 5.2 & X & X & X & 1.0 \\ \hline
	\end{tabular}
    \label{thermal_paras}
    \end{center}
\end{table*}

\subsection{RHESSI imaging at various energies in the thermal range}\label{imaging}

The flare was imaged using {\em RHESSI} detectors 3-7 over five energy bands
(10-11, 11-13, 13-16, 16-20 and 20-25 keV), where the coronal X-ray emission dominated (Figure \ref{flare_intro}). Detectors 1, 2, 8 and 9 were not used. Detector 1 showed no modulation (the source was over-resolved with significant noise), detector 2 is only sensitive above $\sim 20$ keV, and the resolutions of detectors 8 and 9 (106$\arcsec$ and 183$\arcsec$ respectively) are larger than the image area of 64$\arcsec\times64\arcsec$.
The images were created using two imaging algorithms: CLEAN \citep{1974A&AS...15..417H, 2002SoPh..210...61H}
and Visibility Forward Fitting (Vis\_FwdFit) \citep{2007SoPh..240..241S} using a 1 arcsecond pixel size. CLEAN images show
a loop-like structure visible up to the 20-25~keV bin. Such a loop-like structure is well suited for fitting a simple
curved elliptical Gaussian fit, using Vis\_FwdFit\footnote{http://hesperia.gsfc.nasa.gov/rhessi3/software/imaging-software/vis-fwdfit/index.html}, to {\em RHESSI's}
X-ray visibilities, so that the location and the characteristic sizes of the X-ray source can be found. The Vis\_FwdFit $\chi^{2}$ values determine whether the parameters provided by the fit are acceptable.
We note that Vis\_FwdFit
actually provides the fitted curve centre of mass and this is not equivalent to the loop top position
which is required for the study. Instead, the image peak is found using the {\em Solar SoftWare (SSW)} routine $parapeak.pro$. This program estimates the peak position by performing a parabolic fit to the image data. An error is found using the standard deviation of the parabolic fit. The CLEAN image peak is also found using the same method so that the positions provided by each algorithm can be compared. Figure \ref{images} shows a CLEAN image for an energy range of 13-16 keV.
Vis\_FwdFit $50\%$ contours at 10-11 keV, 13-16 keV and 20-25 keV are displayed. Figure \ref{images}
also shows a footpoint source at 60-80 keV.

\subsection{The relationship between X-ray energy and radial distance}

\begin{figure*}
\centering
\includegraphics[width=0.9\linewidth]{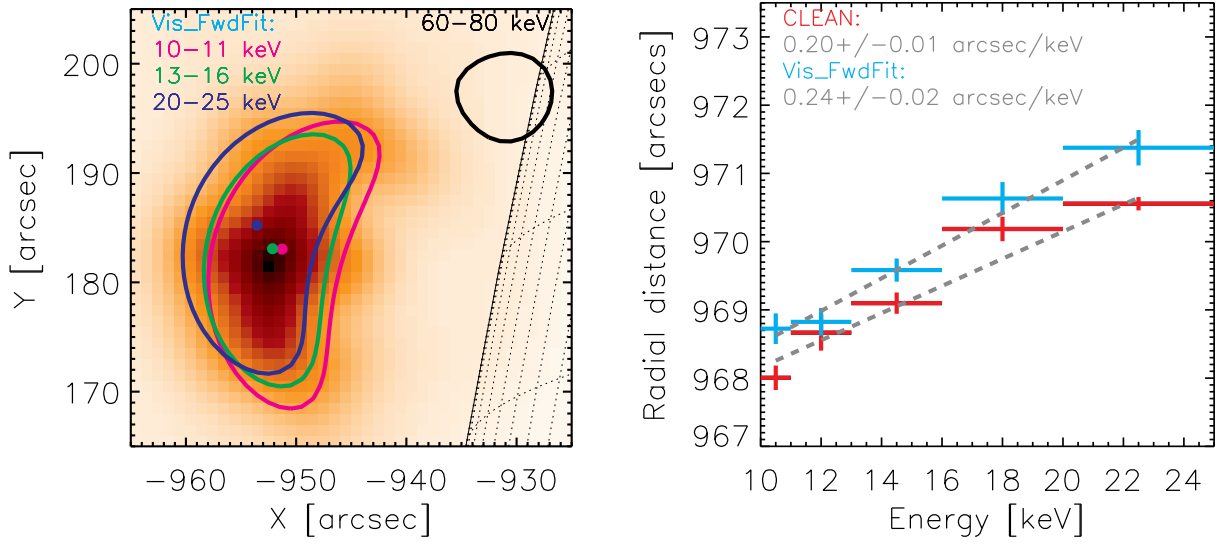}
\caption{{\it Left:} CLEAN image for a time interval from 02:08 to 02:10~UT at 13-16 keV with a CLEAN 60-80 keV 50\% contour (black). Vis\_FwdFit contours at 50\% of the maximum and coronal source positions (coloured dots) are displayed for three energy ranges (10-11 keV (pink), 13-16 keV (green) and 20-25 keV (blue)) showing the increasing altitude of the coronal source location with energy.
{\it Right:} The peak radial position $R$ plotted against X-ray energy $\epsilon$, for both the CLEAN (red) and Vis\_FwdFit (blue) algorithms. The gradient $\alpha=dR/d\epsilon\pm{\rm error}$ found from each linear fit is shown (grey dashed lines) on the graph.}
\label{images}
\end{figure*}

Using the CLEAN and Vis\_FwdFit image peak positions $(x,y)$ provided by $parapeak.pro$,
we calculated the radial distances
$R=\sqrt{x^{2} + y^{2}}$ (measured from the solar disk centre) for each algorithm at each energy range.
An error for $R$ is found from error propagation. Figure \ref{images} (right) shows that the calculated $R$ errors are small for this flare
(approximately half an arcsecond or less) at the chosen energy ranges and time, and
that the data from two imaging algorithms give a similar trend - the higher the energy, the higher the location of the peak.
Both results can be well-fitted by a straight line, so that the observed relationship between radial distance $R$
and energy $\epsilon$ is approximated by
\begin{equation}\label{eq:1}
\frac{dR(\epsilon)}{d\epsilon}=\alpha\,,\;\;\;\,\,\,\mbox{[arcsec keV$^{-1}$]}
\end{equation}
where $\alpha$ is a constant gradient.
\begin{figure}
\centering
\captionsetup{width=1.0\linewidth}\hspace{-1.5cm}
\includegraphics[width=10.2cm]{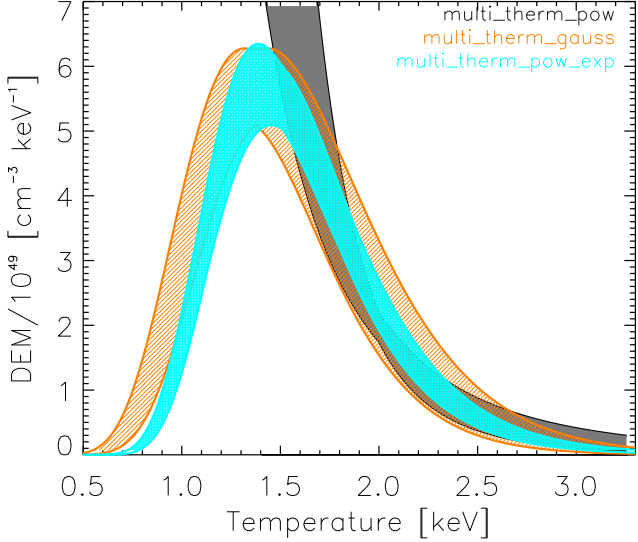}
\caption{Differential emission measures, $DEM(T)$, obtained for the various multi-thermal fits (Figure \ref{fig:spectra}). The shaded areas indicate the estimated confidence intervals. The form of all $DEM(T)s$ are similar between $\sim$1.6 and 3.0 keV, independent of the chosen model $DEM(T)$.}
\label{fig:dems}
\end{figure}
Both algorithms give a similar gradient ${dR}/{d\epsilon}$ equal to $0.24\pm0.02$ arcsec/keV
for Vis\_FwdFit and $0.20\pm0.01$ arcsec/keV for CLEAN, which is important for the analysis. Comparing both algorithms, the absolute values of $R(\epsilon)$ are shifted by $\pm 0\arcsec.3-0\arcsec.4$, with the CLEAN values at a lower height than the Vis\_FwdFit values. This is due to differences in the way each algorithm reconstructs an image. The CLEAN algorithm creates an image by finding point sources and convolving them with the instrumental point spread function (PSF). The image peak will simply correspond to the brightest point smoothed out by the PSF. Since Vis\_FwdFit fits a chosen distribution such as a Gaussian to the X-ray visibilities, the peak of a given source is determined by the fitted distribution to the data. Hence, we should not expect both algorithms to produce the X-ray source peaks in exactly the same locations. For the purposes of the study, Vis\_FwdFit is a more suitable algorithm. However, the systematic difference between both algorithms is very small (sub-arcsecond), regardless of the fact they both create the image in a completely different way. Basically, the CLEAN algorithm is used to provide additional confidence in the results. Other factors, such as the {\em RHESSI} PSF should not alter the locations of the centroids at different energies. The only variable with energy is the transmission of the material overlying the detectors \textemdash\, the attenuators, thermal blankets, and cryostat cover.  While the overall absolute X-ray attenuation of this material is not so well known, its variation with energy is very well known and hence should not produce any systematic effect of the altitude estimates.

\subsection{The vertical extent of the X-ray sources at each energy}

\begin{figure}
\centering
\includegraphics[width=0.9\linewidth]{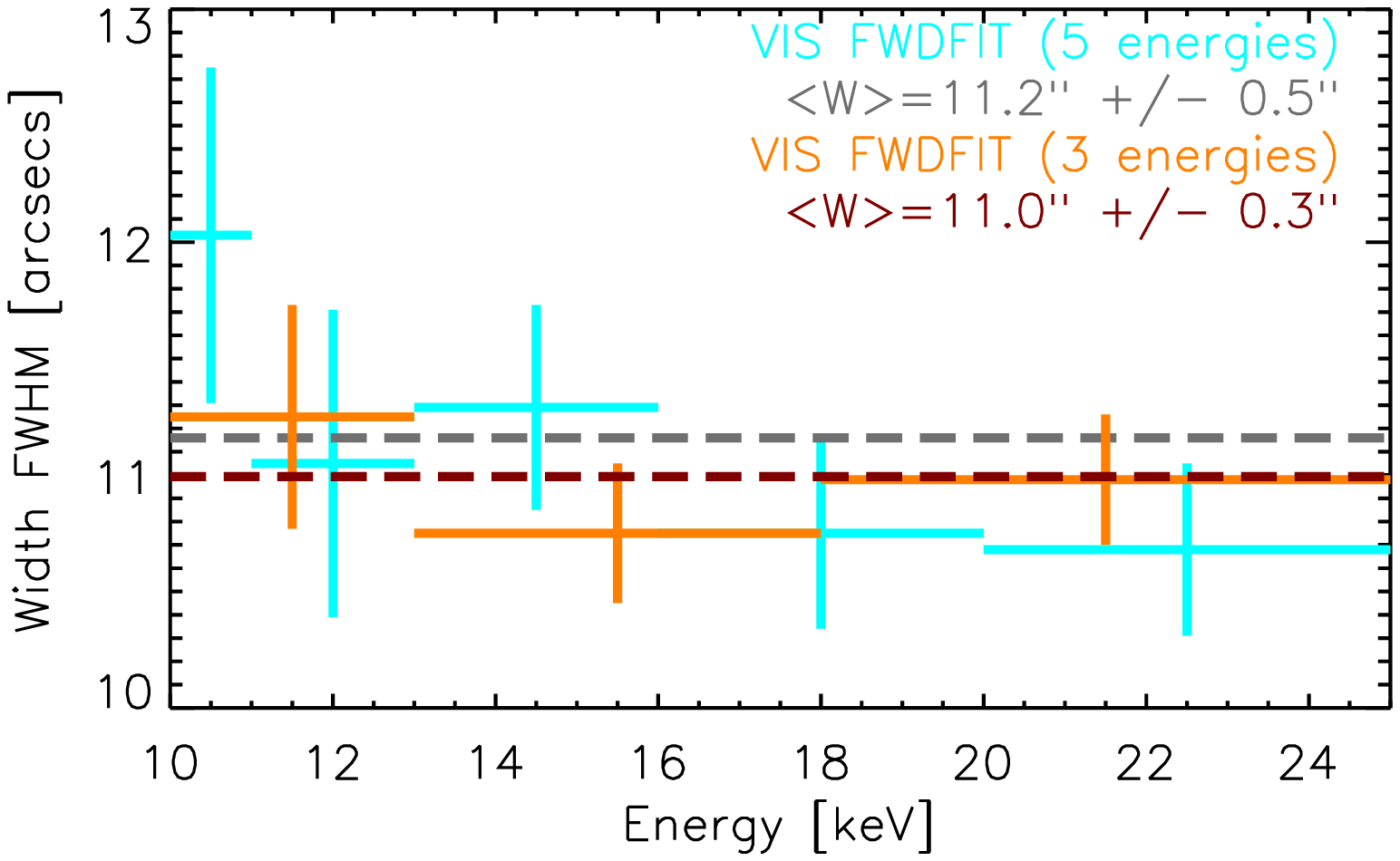}
\includegraphics[width=0.9\linewidth]{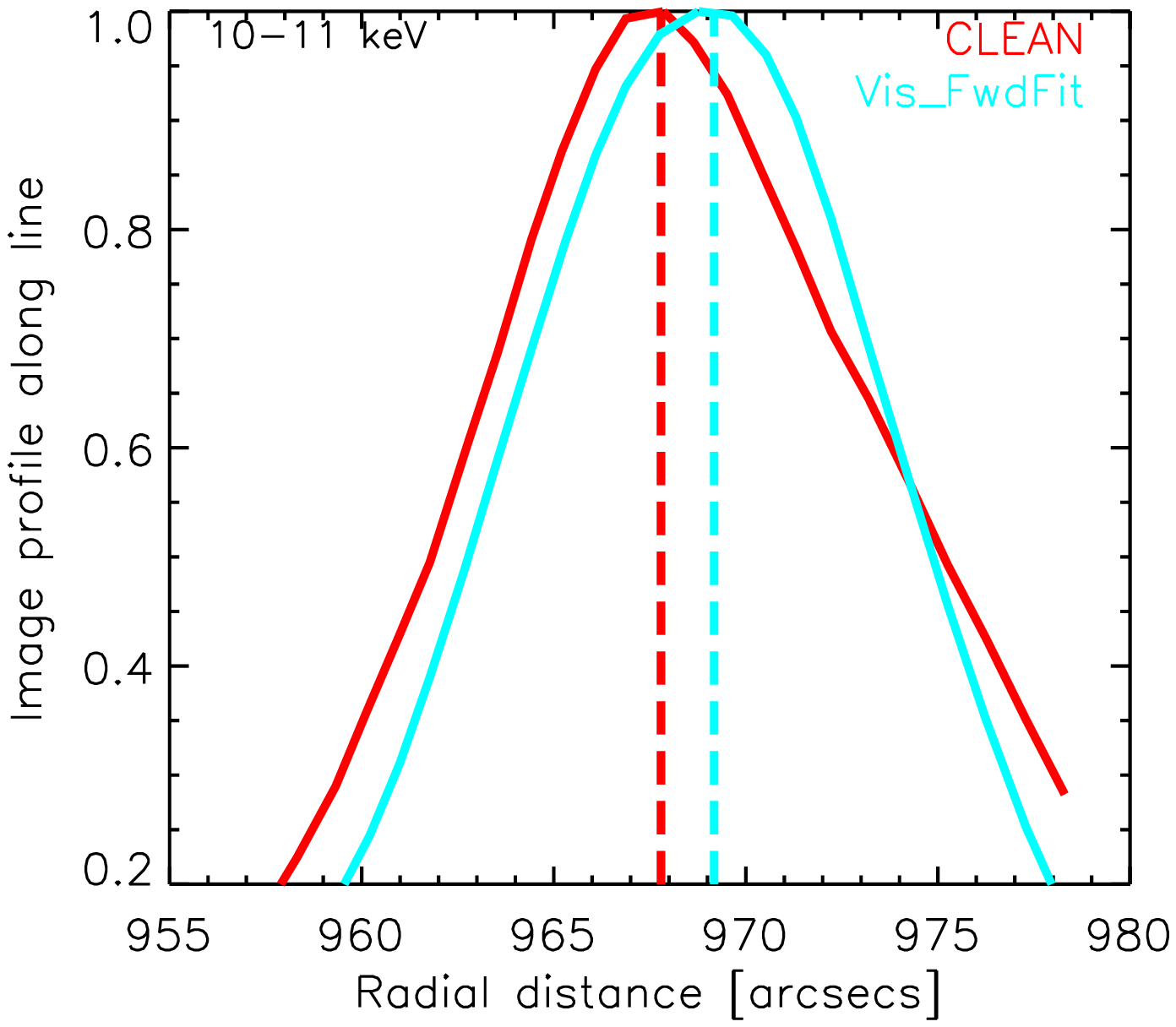}
\vspace{-1cm}
\caption{ {\it Top:} The loop width FWHM $W$ found from Vis\_FwdFit over the same energy ranges as for the $(x,y)$ peak positions shown in Figure \ref{images}. The average width or vertical size of an X-ray source in the radial direction $R$ is $\left<W\right>=11\arcsec.2\pm0\arcsec.5$. The width was also found using three larger energy bins of 10-12.5 keV, 12.5-18 keV and 18-25 keV, increasing the number of counts and hence reducing the uncertainty. This gives an average width of $\left<W\right>=11\arcsec.0\pm0\arcsec.3$.
{\it Bottom:} The CLEAN (red) and Vis\_FwdFit (blue) image profiles along a line perpendicular to the solar surface and through the centroid position of the coronal source, for the energy bin of 10-11 keV. Both curves are divided by the maximum value for comparison. Vis\_FwdFit (blue) fits a Gaussian distribution to the X-ray visibilities, and although the real X-ray intensity profile may deviate from a Gaussian form assumed by Vis\_FwdFit (and the CLEAN PSF), the low $\chi^{2}$ values of the fit tell us a Gaussian is an adequate approximation for the study.}
\label{images2}
\end{figure}

Following \citet{2013ApJ...766...75J}, we determine the characteristic sizes of the coronal sources, using Vis\_FwdFit. The loop width Full Width at Half Maximum (FWHM), $W$, of each X-ray source is found for every energy band (Figure \ref{images2}).
The loop width is defined as the X-ray source FWHM in the radial direction, measured at the loop top, perpendicular to the axis of the curved elliptical Gaussian.
As shown in Figure \ref{images2}, at energies of 10 to 25~keV, the X-ray source width $W$ lies
between $\sim$10$\arcsec$.5 and 12$\arcsec$ and appears to decrease with increasing energy.
The mean X-ray source width is $\left<W\right>=11\arcsec.2\pm0\arcsec.5$.
$W$ is also found
from Vis\_FwdFit using three
wider energy bins of 10-12.5 keV, 12.5-18 keV and 18-25 keV to increase the number of counts per band, with $W$ appearing to remain constant with energy, at least within the errors.
This gives an average width $\left<W\right>=11\arcsec.0\pm0\arcsec.3$.
We note that the width of $\sim11\arcsec$ of each X-ray source is much larger than the observed energy dependent shift in altitude of only $\sim3\arcsec$.

\section{The peak of X-ray emission and DEM(T)}\label{equat}

From {\em RHESSI} observations (see Figure \ref{images}), we found that the flare coronal X-ray source
altitude increases with energy, according to the approximate linear relation (\ref{eq:1}).
Hence
\begin{equation}\label{r_ep}
\frac{dz}{d\epsilon}=\alpha\; ,
\end{equation}
where $z$ is the height above the solar limb.

Our task is to find an analytical model that can produce the observed $z$ versus $\epsilon$ trend and relate such changes
to a temperature $T(z)$ structure and a number density $n(z)$ structure in the flaring coronal region.
To do this, firstly, let us derive a relationship between the altitude $z$ of the X-ray source and energy $\epsilon$.

Since the spectroscopy results for the flare show that the spectrum in the range $\sim 6$ to $25$ keV
can be well-fitted by multi-thermal models (Figure \ref{fig:spectra}),
we will assume that all the coronal source X-rays in this range are emitted as multi-thermal bremsstrahlung. This assumption is supported
by the fact that the thermal emission dominates the non-thermal emission in this range.
The spectroscopy results (Figure \ref{fig:spectra}) show that the thermal and non-thermal components are equal at $\sim 28$~keV with the thermal component being increasingly dominant at lower energies.  Consequently, we will assume that all the emission in the 6 to 25~keV energy range is multi-thermal bremsstrahlung from the flaring coronal source.

Consider an emitting volume $dV=Adz$, where $A$ is the X-ray source area perpendicular to the radial direction (altitude)
and image plane, as shown in Figure (\ref{fig:loop_pic}). The photon flux emitted as thermal bremsstrahlung
per unit energy $\epsilon$ per unit height $z$ from a plasma characterised by a temperature $T$ [keV] and
number density $n$ [cm$^{-3}$] is given by e.g. \citet[][]{1988psf..book.....T}
\begin{equation}\label{I1}
J(\epsilon, z)\propto\frac{n^{2}(z) A(z)}{\epsilon \sqrt{T(z)}}\exp{\left(-\frac{\epsilon}{T(z)}\right)}\;,
\end{equation}
where $J(\epsilon, z)$ is the photon flux per unit of height [photons s$^{-1}$ cm$^{-2}$ keV$^{-1}$ cm$^{-1}$].

The total photon flux [photons s$^{-1}$ cm$^{-2}$ keV$^{-1}$] integrated over all $z$ is then given by,
 \begin{equation}\label{I2}
I(\epsilon)=\int_{0}^{\infty}J(\epsilon, z)dz.
\end{equation}

At any observed energy $\epsilon$, X-ray flux $J(\epsilon, z)$ should have maximum at $z(\epsilon)$ determined by the derivative
\begin{equation}\label{I4}
\frac{\partial J}{\partial z}\propto\frac{\partial }{\partial z}\left[\frac{n^{2}A}{\epsilon\sqrt{T}}\exp{\left(-\frac{\epsilon}{T}\right)}\right]=0.
\end{equation}
Re-arranging Equation (\ref{I4}) for $\epsilon (z)$ one finds,
\begin{equation}\label{I5diff}
\epsilon(z)=\frac{T(z)}{2}
  -\frac{d(n^{2}A)}{dz}\frac{dz}{dT}\frac{T^2}{n^{2}A}.
\end{equation}
and we can write:
\begin{equation}\label{I5}
\epsilon(z)=\frac{T(z)}{2}
  -\frac{d(n^{2}A)}{dT}\frac{T^2}{n^{2}A}.
\end{equation}

The plasma density $n(z)$ and temperature $T(z)$ also determine the differential emission measure
\[
DEM(T)=n^{2}\frac{dV}{dT}.
\]
Using the definition of $DEM(T)$, one can write
\begin{equation}\label{dem_model}
DEM(T)=\left.n^{2}\frac{dV}{dT}\right|_{z=z(T)}
=\left.n^{2}A\frac{dz}{dT}\right|_{z=z(T)}
\end{equation}
where $z=z(T)$ is the height at a temperature $T$.

Both $DEM(T)$ and the peak position of the X-ray flux at a given energy $\epsilon(z)$ are uniquely determined by $T(z)$ and $n^{2}(z)A(z)$.

\subsection{Isothermal plasma}\label{case1}

Let us first assume a simple case. For an isothermal plasma $dT/dz=0$ and hence $d(n^2A)/dT=0$, Equation (\ref{I5}) becomes
\begin{equation}\label{I_iso}
\epsilon_{0}=\frac{T}{2}=\frac{T_{0}}{2}
\end{equation}
over all $z$, where $T=T_{0}$ is a constant. Therefore, an isothermal plasma {\it cannot} account for the observed changes in peak X-ray energy with height $z$.
Importantly, this shows that \textit{an isothermal plasma in the flaring coronal source} is ruled out by imaging data, even though an isothermal bremsstrahlung function can reasonably fit the solar flare spectrum found using {\em RHESSI} (see Figure \ref{fig:spectra}).

\subsection{Height-dependent temperature with constant $n^{2}A$}\label{case2}

If $n^{2}A$ is constant with height $z$, i.e. $d(n^2A)/dz=0$, then Equation (\ref{I5diff}) yields
\begin{equation}\label{I_A}
\epsilon_{0}=\frac{T(z)}{2}.
\end{equation}
Since the observed X-ray energies are $10$ to $25$~keV, the temperatures $T(z)$ in a constant $n^{2}A$ plasma would have to be of the order $20$ to $50$~keV, i.e. $230$ to $580$ MK. From the {\em RHESSI\,} spectrum, we see that the majority of the emitted thermal X-rays are emitted at lower energies corresponding to lower temperature sources of $1$ to $3$~keV, and hence a constant $n^{2}A$ is unlikely.

\subsection{Height-dependent $T(z)$ and $n^{2}(z)A(z)$}\label{case5}
In Equation (\ref{I5}), if the gradient $d(n^{2}A)/dT\neq0$, then $d(n^2A)/dT=(d(n^2A)/dz)(dz/dT)$ must be negative for energy $\epsilon$ to be positive, {\it or} the second term must be smaller than the first term on the RHS of Equation (\ref{I5}). Since we observe higher energy X-ray sources above lower energy X-ray sources, the temperature must increase with altitude, so that $dT/dz>0$. Hence, in this scenario $d(n^2A)/dz<0$ for $d(n^2A)/dT<0$. However, $d(n^2A)/dz$ does not have to be negative over all $z$, only over a portion of $z$ where we see the X-ray flux peak positions at a given energy $\epsilon$. Actually, if $d(n^2A)/dz$ is negative over the entire region, then the X-ray flux peak positions may appear at a lower altitude than suggested by a given observation, but this will further investigated via modelling. It is more difficult to determine the form of $\epsilon$ versus $z$ for the case where $d(n^2A)/dz$ is positive, and the second term is smaller than the first term. Hence modelling is required to further determine the forms of $T(z)$ and $n^{2}(z)A(z)$ that can account for both {\em RHESSI} imaging and spectroscopic observations.

\section{Modelling the vertical temperature and number density distributions}\label{models}

Using X-ray imaging and spectroscopy data as a guide,
we develop a model of $T(z)$ and $n^{2}(z)A(z)$, which can \textit{simultaneously} explain flare X-ray imaging
and spectroscopic observations.

\begin{figure*}
\centering
\includegraphics[width=0.9\linewidth]{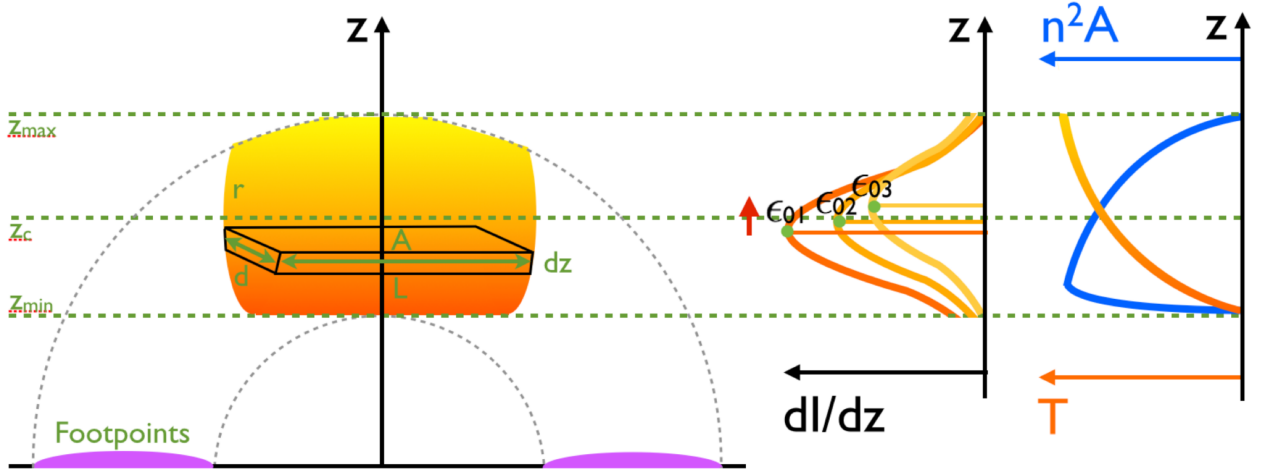}
\caption{{\it Left:} Sketch of the coronal X-ray source geometry, as seen by {\em RHESSI}. The lower temperature plasma is located below the higher temperature plasma.
A small volume of emitting plasma $dV=A\,dz=Ld\,dz$ varies with altitude $z$.
{\it Right:} Cartoon of the X-ray distribution versus $z$ at energies $\epsilon_{01}<\epsilon_{02}<\epsilon_{03}$
created by varying temperature $T$ and $n^{2}A$ distributions with altitude $z$.
The peak X-ray positions at a given energy, as viewed by {\em RHESSI}, are related to a $T$ and $n^{2}A$ value at that position.}
\label{fig:loop_pic}
\end{figure*}

Firstly, we note that the density $n(z)$ and the area perpendicular to the image plane $A(z)$ (see Figure \ref{fig:loop_pic} for the geometry)
appear in Equation (\ref{I5}) and Equation (\ref{dem_model}) as a combination $n^{2}(z)A(z)$. Therefore, the observations cannot say
anything further about the individual distributions of $n(z)$ and $A(z)$, only about the combination  $n^{2}(z)A(z)$.
However, we can deduce the form of $n^{2}(z)A(z)$ and the temperature structure $T(z)$ from the observations.

The modelled $n^{2}(z)A(z)$ and $T(z)$ should reproduce the following main characteristics of X-ray emission determined from {\em RHESSI} observations:

\begin{enumerate}

\item{The observed X-ray source widths (using a Gaussian fit) should have a FWHM$\sim 11\arcsec$ with visible X-ray emission over $\sim20\arcsec$ in the radial direction between the energies of 10 to 25 keV.
The minimum visible X-ray emission should be at a height $\sim10\arcsec$ above the limb.}

\item{Altitude $z$ versus peak energy $\epsilon$ should be well-fitted by Equation (\ref{I5}), relating energy, height, $T(z)$ and $n^{2}(z)A(z)$. The gradient $\alpha$ has a value of about $+0.2$ arcsec/keV, which means the height difference
between the peak positions of the minimum and maximum energies of 10 and 25 keV is about $3\arcsec$.}

\item{The $DEM(T)$ derived from modelling should be consistent with the $DEM$ used to fit the spatially integrated {\em RHESSI}
spectrum (see Figure \ref{fig:dems}). From {\em RHESSI} spectroscopy, we noted that the OSPEX model $DEM$ functions agree well within the temperature
    range of $\sim$1.6 to 3.0 keV.}

\end{enumerate}

\begin{figure*}
\centering
\includegraphics[width=0.9\linewidth]{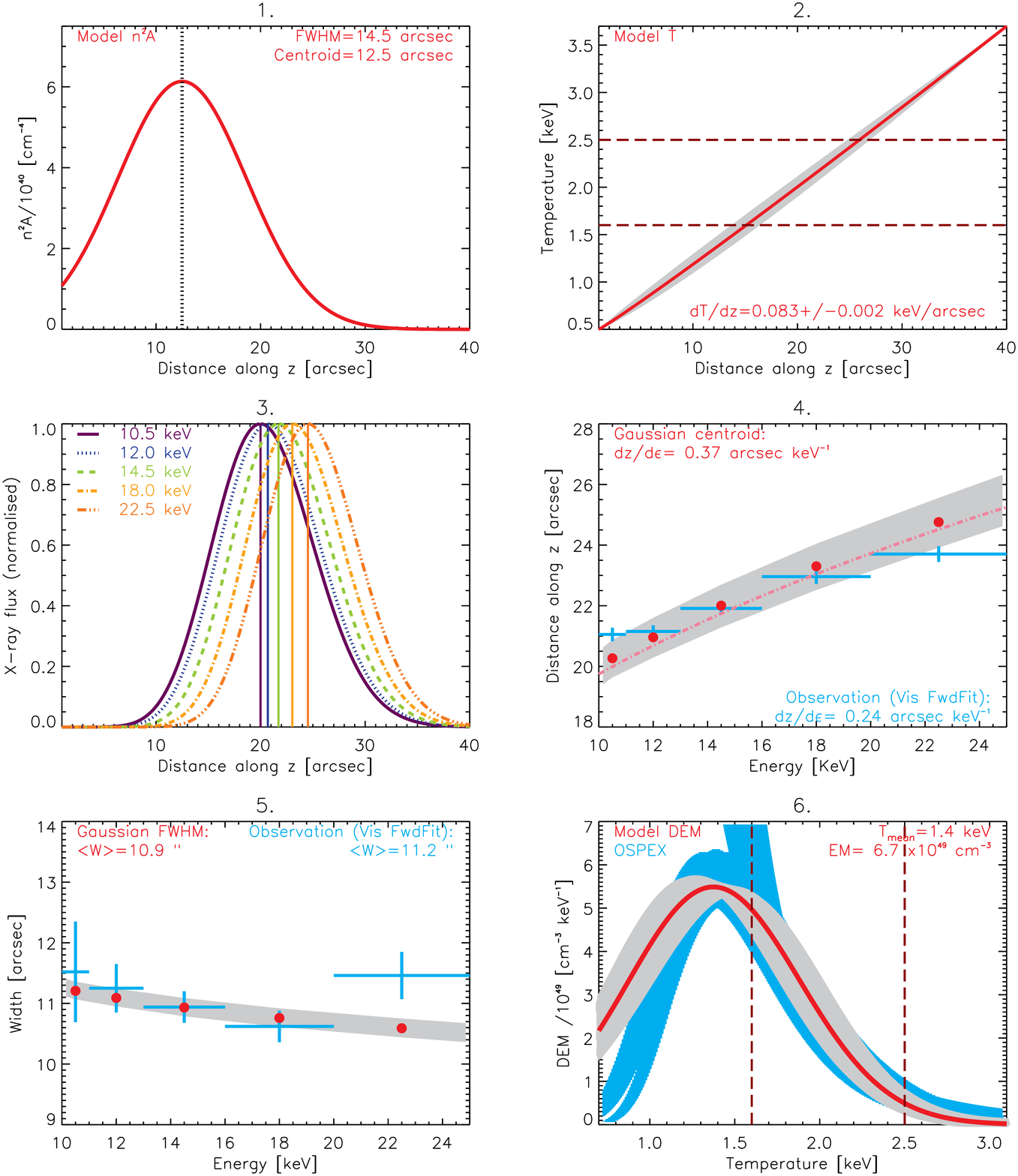}
\caption{The temperature and density-altitude distribution model (see Section \ref{models} for details) with 6 different panels. In all panels, the best model result is represented by the red line and points.
1: $n^{2}A$ vs. height $z$ given by Equation (\ref{naz}).
2: Temperature $T$ vs. $z$ given by Equation (\ref{tz}).
3: X-ray flux at various photon energies $\epsilon$ vs. altitude. The vertical lines
indicate the peak locations.
4: The model X-ray distribution peak position vs. energy (red dots) and Equation (\ref{I5diff}) using the model $T(z)$ and $n^{2}(z)A(z)$ (pink dashed-dotted line).
5: X-ray distribution widths (FWHM) vs. energy.
6: Differential Emission Measure, $DEM(T)$. Panels 4 -  6 show both the results of the {\em RHESSI} data analysis (in blue) and the model predictions (in red).
The grey areas show the range of each parameter that can adequately match the measured values using the range of modelled $T(z)$ shown in Panel 2.}
\label{model_good}
\end{figure*}

The main results are presented in Figure \ref{model_good}, which has six panels showing: (1) $n^{2}(z)A(z)$, (2) $T(z)$, (3) X-ray flux at various photon energies $\epsilon$,
(4) The position of the X-ray distribution peak versus energy,
(5) X-ray distribution widths (FWHM) versus energy, (6) $DEM$ versus temperature.

We note that {\em RHESSI} imaging observations at a given energy $\epsilon$ show that the X-ray profiles
along $z$ can be well-described by a Gaussian distribution (i.e. using Vis\_FwdFit, even if a Gaussian is not the exact, true form of the profile - see Figure \ref{images2}). Hence from Equation (\ref{I1}), the easiest way to produce a Gaussian profile, is to create a $n^{2}(z)A(z)$ profile that is also Gaussian or close to Gaussian along $z$. Hence taking $n^{2}(z)A(z)$ [cm$^{-4}$] as a Gaussian distribution, we have
\begin{equation}\label{naz}
n^{2}(z)A(z)=n^{2}_{0}A_{0}\exp{\left(-\frac{(z-z_{s})^{2}}{2\sigma^{2}}\right)}.
\end{equation}
with centroid $z_{s}$ and FWHM $=2\sqrt{2\ln{2}}\sigma$.
Here, we choose FWHM$=14\arcsec.5$ and $z_{s}=12\arcsec.5$, as measured above the solar X-ray limb, and $n^{2}_{0}A_{0}=6\times 10^{40}$ cm$^{-4}$. All these values are found via trial and error, and they are chosen because they produce the main flare observables. For a constant value of $A$, Equation (\ref{naz}) suggests that the number density would have to fall an order of magnitude or more
with increasing $z$, as can be deduced from Panel 1 in Figure \ref{model_good}.
For the chosen $n^{2}(z)A(z)$, the peaks of the X-ray source at a given energy only appear when $d(n^{2}A)/dz$ becomes negative for $z>z_s$ (to the right of the black dotted line in Figure \ref{model_good}), i.e. X-ray emission appears at lower altitudes but the peaks of the X-ray source at a given energy can only appear for $z > z_{s}$.

The temperature $T(z)$, as function of vertical height $z$, is modelled as a power-law with fixed minimum and maximum
temperatures (Figure \ref{model_good}), so that
\begin{equation}\label{tz}
T(z)=0.5+3.2\left(\frac{z-z_{min}}{z_{max}-z_{min}}\right)^{\gamma}\,\mbox{keV}
\end{equation}
where $0.5$ keV ($\sim 6$~MK) at $z_{min}=1\arcsec$ and $3.7$~keV ($\sim 43$~MK) at $z_{max}=40\arcsec$.
As discussed in Section \ref{case5}, the observations require $T(z)$ increasing with height $z$, so that $dT/dz>0$.
The temperature gradient in Equation (\ref{tz}) is changed by varying the power index $\gamma$.
Figure \ref{model_good} shows the variations of $\gamma$ between 0.95 and 1.15 (grey region). This represents
a range of $T(z)$ that can best reproduce all the {\em RHESSI} observations within uncertainties,
while the red curve represents the $T(z)$ distribution with $\gamma\simeq 1.1$ that best fits all the {\em RHESSI} observations.

Panel 4 in Figure \ref{model_good} shows the resulting X-ray flux against height $z$, for five chosen X-ray energies,
chosen to match the average values of the energy ranges used for {\em RHESSI} imaging observations.  Since $n^{2}A$ was input as a Gaussian,
the resulting $I(z)$ at each energy $\epsilon$ are also close to Gaussian, as suggested by the observations.
The model X-ray peak positions versus energy give a gradient of $dz/d\epsilon\simeq 0.37$ arcsec/keV, which is relatively close to the observed
$dz/d\epsilon=0.24$ arcsec/keV. The best-fit X-ray peak positions also appear at a height $z$ of $\sim20\arcsec$ (at 10 keV),
as suggested by {\em RHESSI} imaging. The average FWHM $\simeq 11\arcsec$ matches the observed average,
and shows that the width decreases with energy as  suggested by the images.

In the temperature range between 1.6 keV and 3.0 keV, all modelled $DEMs$ match well with the $DEM$ models fitted to {\em RHESSI}
spectrum.

The model emission measure $EM$
\begin{equation}\label{em}
EM=\int_{T}DEMdT
\end{equation}
and the average temperature $\left<T\right>$
\begin{equation}\label{Tmean}
\left<T \right>=\frac{\int_{T}T\,DEM\,dT}{\int_{T}DEM\,dT},
\end{equation}
are found to be $EM \sim 6.7\times10^{49}$ cm$^{-3}$ and $\left<T\right>=1.4$ keV ($\sim16$ MK).

Over the temperature range between 1.6 and 2.5 keV, the model $z$ varies from around $\sim15\arcsec$ to $\sim25\arcsec$ (see Figure \ref{model_good} Panel 2).
This leads to a model temperature gradient of $dT/dz\sim0.08$ keV/arcsec. From the model, it is suggested that the form of the $DEM$ at lower temperatures below $1.6$ keV is closer to both the Gaussian and power-law exponential OSPEX models, which could not be confidently determined by spectroscopy alone.

\begin{figure*}
\includegraphics[width=0.33\linewidth]{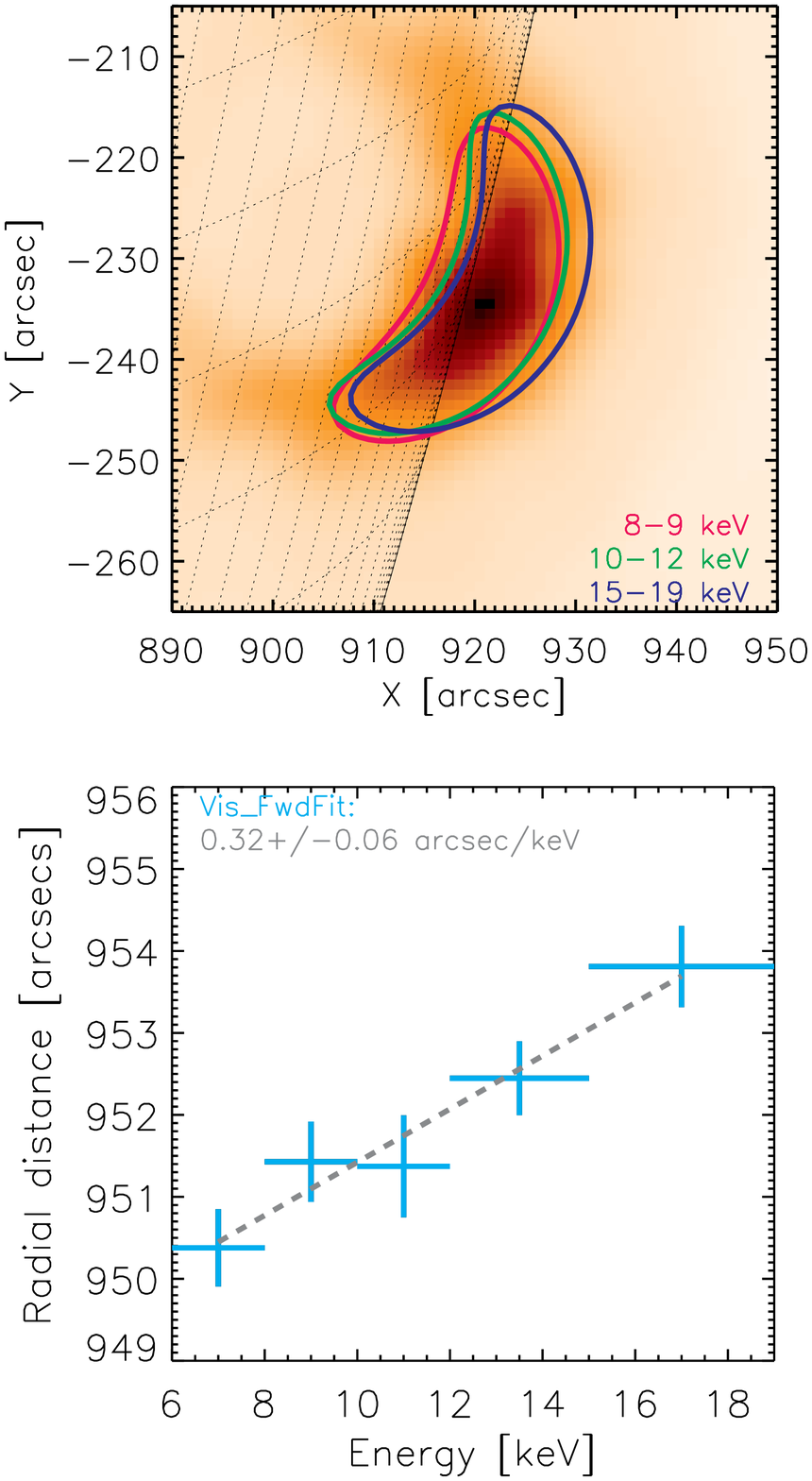}
\includegraphics[width=0.33\linewidth]{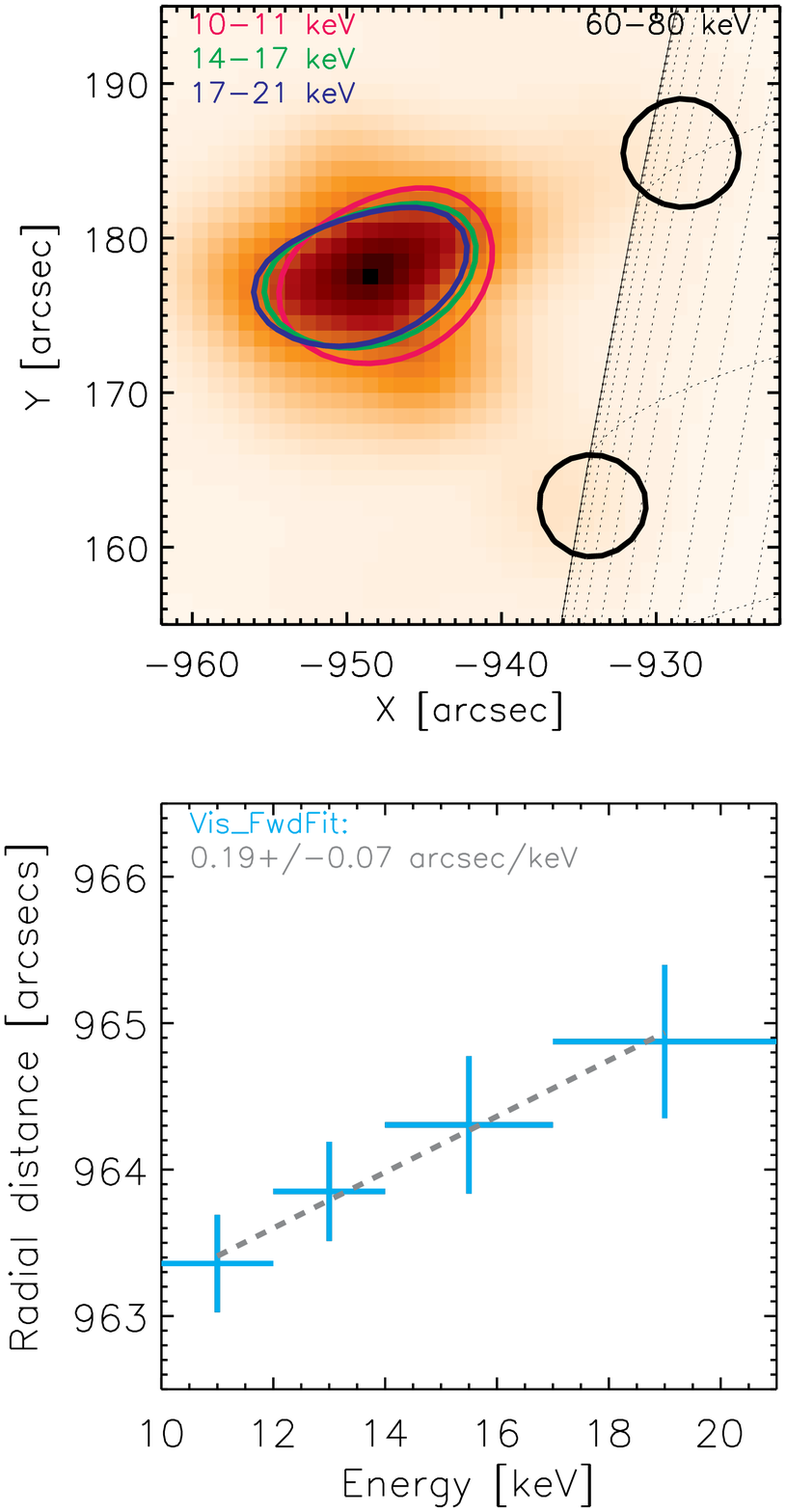}
\includegraphics[width=0.33\linewidth]{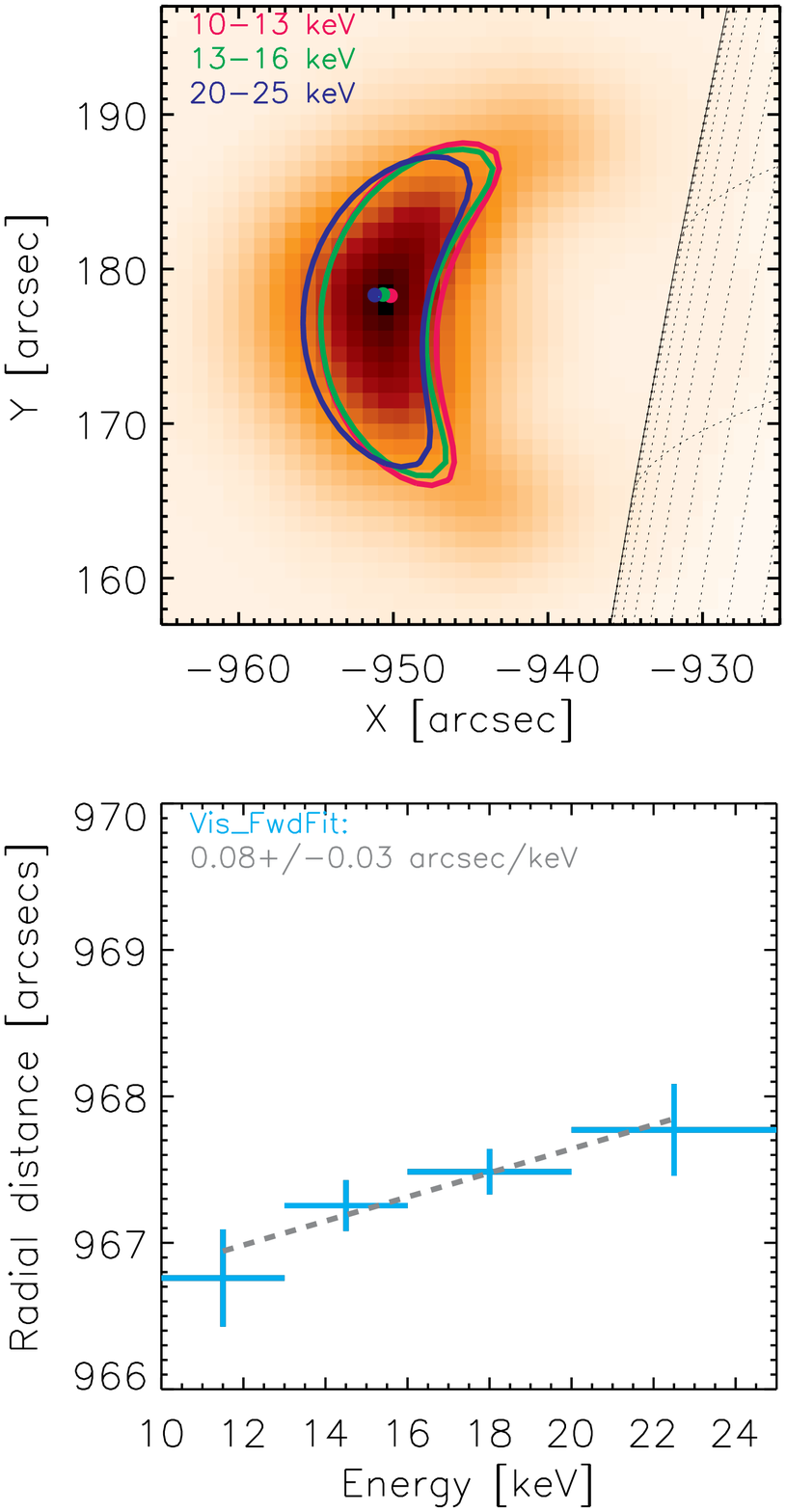}
\caption{Images of coronal X-ray sources in different energy bands, for two additional flares: SOL2005-08-23T14:32 from 14:30 to 14:33 UT (left) and SOL2013-05-13T16:05 from 16:02 to 16:04 UT (middle). Both flares are imaged during the time of peak X-ray flux, and the source altitude increases with X-ray photon energy in these two flares. This trend also occurs for the analysed flare SOL2013-05-13T02:12 at different times (right). Here the imaging results are shown for the time of 02:04 to 02:06 UT during the rise phase of the flare, showing that the gradient $dR/d\epsilon$ varies with time.}
\label{all_flares_1}
\end{figure*}

In Figure \ref{all_flares_1} (right most panel), we also show that such a trend is common at different times during the studied flare, with the gradient $dR/d\epsilon$ varying at different times. However, changes in time are beyond the scope of this paper. The temporal changes are the subject of ongoing studies. In Figure \ref{all_flares_1} (middle and right most panels), we also show that the coronal X-ray source altitude increases with energy for two other flares studied at a single time range close to the peak flux of each flare.

\section{Summary}

A detailed analysis of solar flare SOL2013-05-13T02:12, during a single time interval,
was performed using {\em RHESSI}
imaging and spectroscopy observations. We investigated quantitatively the increase in altitude of the flare
coronal X-ray source with energy at a single time interval during the flare impulsive phase,
a trend that has been noted previously \citep[e.g.][]{2013ApJ...766...75J}.

The {\em RHESSI} imaging analysis shows that the peaks of X-ray emission in the coronal X-ray source
are located close together over a small distance of $3\arcsec$ between the X-ray energies of 10 and 25 keV.
The increase in source height with photon energy can be well-fitted by a linear function
with a gradient of $dz/d\epsilon\simeq 0.2$ arcsec/keV. {\em RHESSI} spectroscopy showed that a number
of different thermal models (both isothermal and multi-thermal) can fit the spatially integrated X-ray spectrum adequately well.
At the same time, the analysis of the flare data shows that the emitting flare plasma in the corona {\it cannot\,}
be isothermal, as an isothermal plasma is unable to account for the observed changes in altitude
with X-ray energy, as shown by {\em RHESSI} imaging.
This is an important result for deriving the properties of hot X-ray emitting plasma in the flaring solar corona,
which is often performed with an isothermal model.

Further, our study shows how {\em RHESSI} imaging can be used to constrain the properties of flaring plasma beyond what is possible using X-ray spectroscopy alone.
Coronal X-ray emission in the range of 10 to 25 keV can be well explained using model distributions in
temperature, number density and area, that vary with altitude. Modelling can adequately
explain all the main observations: radial position (gradient $dz/d\epsilon$ and approximate
height above the X-ray limb), vertical source size, and differential emission measure.
For our model, the temperature and $n^{2}A$ gradients ($dT/dz$ and $dn^{2}A/dz$) must have
opposite signs and the $n^{2}A$ distribution must decrease over a portion of the region with increasing altitude $z$, in order to produce the {\em RHESSI} imaging results.
The modelling suggested that $dT/dz$ between 1.6 and 2.5 keV should be $\sim0.08$ keV/arcsec. Since this gradient occurs over most of the observable region ($\sim15\arcsec$), it may be used in future analysis to distinguish between different cooling and heating processes occurring within the flaring coronal X-ray source.
A constant area $A$ suggests that the number density should decrease over an order of magnitude with altitude, although $A$ might also vary with $z$, as suggested by observations of X-ray source length and width variations.
Again, such a prediction should help constrain the processes occurring in the coronal region of solar flares.

We have analysed a number of other limb flares (see Figure \ref{all_flares_1} for some examples) with bright coronal X-ray sources, and they all show the source altitude increasing with increasing X-ray energy.
This suggests that the observations
presented here are common for a majority of coronal sources, and hence \textit{vertical} temperature
and density gradients with rather broad multi-thermal $DEM(T)s$ are common.

One notable limitation of {\em RHESSI} is the insensitivity to plasma at temperatures below $\sim 1$~keV. Different imaging
instruments should be used to extend the range of temperatures \citep[e.g.][]{2013ApJ...779..107B}. We plan to better constrain the electron number density and temperature
distributions of a flaring coronal region, supplementing existing {\em RHESSI} data with EUV imaging and spectroscopy data.
This will allow us to extend our imaging approach to temperatures down to a few MK.

In conclusion, we have demonstrated how imaging should be used to improve spectral analysis, allowing the shape of the $DEM(T)$ to be better deduced, with X-ray imaging providing a relatively simple guidance to the $n^2(z)A(z)$ and temperature
 $T(z)$ profiles within a flaring coronal X-ray source.

\begin{acknowledgements}
NLSJ was funded by a STFC STEP award. EPK gratefully acknowledges the financial support by the STFC Consolidated Grant.
\end{acknowledgements}

\bibliographystyle{aa}

\end{document}